\documentclass[10pt]{article}

\usepackage[fleqn]{amsmath}   
\usepackage{amssymb}
\usepackage{graphicx}
\usepackage{color} 
\usepackage{amsbsy}
\usepackage{setspace} 
\usepackage{booktabs}
\usepackage{authblk}
\usepackage{url}

\usepackage[labelfont=bf,labelsep=period,justification=raggedright]{caption}

\usepackage{natbib}
\makeatletter
\renewcommand{\@biblabel}[1]{\quad#1.}
\makeatother

\newcommand{\oh}{{\includegraphics[height=1.5ex]{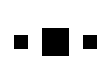}}}
\newcommand{\ov}{{\includegraphics[height=1.5ex]{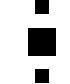}}}

\begin{document}

\title{Texture features for the reproduction of the perceptual organization of sound.}
\author[1,3,*]{Ronald A.J. van Elburg}
\author[1,2,3]{Tjeerd C. Andringa}
\affil[1]{Institute of Artificial Intelligence, Faculty of Mathematics and Natural Sciences, University of Groningen, Groningen, The Netherlands}
\affil[2]{University College Groningen, University of Groningen, Groningen, The Netherlands}
\affil[3]{SoundAppraisal BV, Groningen, The Netherlands}
\affil[*]{Correspondence: RonaldAJ@vanElburg.eu}

\maketitle
\date{}
\begin{abstract}
Human categorization of sound seems predominantly based on sound source properties. To estimate these source properties we propose a novel sound analysis method, which separates sound into different sonic textures: tones, pulses, and broadband noises. The audible presence of tones or pulses corresponds to more extended cochleagram patterns than can be expected on the basis of correlations introduced by the gammachirp filterbank alone. We design tract features to respond to these extended patterns, and use these to identify areas of the time-frequency plane as tonal, pulsal, and noisy. Where an area is marked as noisy if it is neither tonal nor pulsal. To investigate whether a similar separation indeed underlies human perceptual organization we introduce tract based descriptors: tonality, pulsality, and noisiness. These descriptors keep track of either the total energy or the cochleagram area marked as respectively tonal, pulsal, and noisy. Energy based tonality and pulsality is strongly correlated with the first perceptual dimension of human subjects, while energy based noisiness correlates moderately with the second perceptual dimension. We conclude that harmonic, impact and continuous process sounds can be largely separated with energy based tonality, pulsality and noisiness. \\
\textbf{{Keywords:} Sound Perception, Computational Auditory Scene Analysis, Sound Texture, source properties, time-frequency analysis  } 
\end{abstract}

\section{Introduction}
\subsection{Sound Perception}
In contrast to vision, where the source of the light is less important than the object that reflected it last, audition is geared towards the sound source rather than the objects that reflected the sound ~\cite[]{Andringa2010}. This suggests that production physics plays an essential role in audition. In addition, human sound perception is by and large independent of the environment in which the sound is produced. Typically, the production environment contributes reverberation effects as causal filtering and uncorrelated sources as additive "noise". According to \cite{Nabelek_1982}: "the presence of [additive] noise can be easily noticed. The presence of moderate reverberation is not apparent. It takes a trained ear, or paired comparisons of speech samples recorded without and with reverberation to detect the presence of reverberation." This suggests that while listeners may not be very sensitive to transmission effects, they are highly sensitive to production physics. 

In fact production physics may underlie perceptual organization and similarity. \cite{Gaver_1993} notes that all sounds result from interacting materials and he proposes to subdivide sounds by their physical origin: sounds of vibrating solids, aerodynamic sounds, and liquid sounds. While this leads to a suitable taxonomy for some purposes, it cannot be related uniquely to perception and signal processing because the impact sound of two solids may resemble the sound of a gas explosion. Alternatively, source physics and signal processing can be related uniquely to perception by focusing on impulse responses and resonances as factors shaping sound. Sharp resonances characteristically produce tones or harmonic complexes locked to a fundamental frequency contour. Both lead to horizontal structures in the time-frequency plane: structures extended in the time direction. Impulse responses of strongly damped systems lead to short pulse-like sounds with associated  vertical structures in the time-frequency plane: structures extended in the frequency direction. The impulse response of a weakly damped resonator, for example a tuning-fork, results in a sharp onset, a pulse, followed by a damped tone. The associated structure in the time-frequency plane is a "vertical" onset followed by a "horizontal" ringing out. 

While tones and pulses show-up as narrow structures in the time-frequency plane, broadband noises appear as broad excitations, extended in both time and frequency, with a fluctuating and, typically, never-repeating fine-structure. Broadband noises can be interpreted as a combination of many uncorrelated sources. The sound of flushing a toilet represents such a multitude of uncorrelated "impacts" and "resonances" --- at multiple spatial and temporal scales --- that the sound it produces becomes broadband. Alternatively, light rain will be dominated by the pulses of individual drops hitting objects, while heavy rain will shift towards broadband noise. Similarly individual speakers will be dominated by tonal contributions, while the babble of many speakers shifts towards broadband noise. In the limit of infinite contributions both rain and babble resemble a broadband spectrum.

We assume that nearly every point of the time-frequency plane of a natural sound will be dominated by energy contributions that are either tonal, pulsal, or noisy (we mean ‘broadband noise’ when we use the term ‘noise’). To demonstrate the relevance of this separation to human audition we compare the (energy weighted) fraction of pulsal, tonal and noisy contributions with the perceptual dimensions identified by a perceptual similarity experiment of environmental sounds by \cite{Gygi_2007}. This experiment provides the human reference data we need to determine whether our separation of sound into tones, pulses, and broadband noises approximates perceptual distance judgments.

\subsection{Finding Structure Attributable to Sound Sources}

To produce a time-frequency representation for sound we need to apply a set of (digital) filters parametrized by frequency and time to the incoming signal. Formally speaking, because there are no true repetitions, it would be more correct to speak of time-scale representations in the following, but this is against the convention used in CASA. We will therefore use the commonly used term time-frequency. 

The resolution in a time-frequency plane is intrinsically limited by the Heisenberg uncertainty relations. The highest possible resolution is obtained with filter functions that show a minimal spread in the time-frequency plane. However, minimal spread is only defined with respect to the exact choice of time and frequency operators for which the spread is minimized. It is in fact possible to make different time-frequency representations leading to different minimal spread filters~\cite[]{Cohen1993, Irino1995, Reimann2011}. The popular gammachirp filterbank, a linear model for the basilar membrane, uses a frequency dependent minimal spread representation minimizing the spread in different but related time and frequency operators at each frequency~\cite[]{Irino1995}. The use of a frequency dependent wavelet representation allowed \cite[]{Irino1999} to fit the representation to electro-physiological data from the cochlea. Because of this physiological plausibility we use the gammachirp representation in our comparison of tract features with human psychophysical data. However, after recalibration our analyses and code can be applied to time-frequency or time-scale representations produced by filterbanks related to other transformation pairs as well.

In addition to producing a two dimensional representation from a one dimensional input signal, each filterbank introduces its own correlations in the time-frequency plane. To distinguish between filter-induced structures and input or sound source related structure, we propose a statistical approach. In our approach, time-frequency plane structures that are unlikely to be introduced by the filterbank are attributed to structure in the input and considered indicative of source physics. Structures likely to be introduced by the filters are attributed to unstructured input. We make several choices to make this idea operational: first regarding the type of structures we try to establish in the output, and secondly regarding threshold parameters which we use to decide whether or not a value is indicative of source structure. Consequently we develop a set of measures that indicate whether structure in the filterbank output should be attributed to the filterbank or are more likely to reflect source imposed structure.

To characterize the filter induced structure we apply the filters to an unstructured reference input signal. We need to decide what kind of input signal we will call unstructured. Gaussian white noise signals seem a natural choice because a discretized signal $s(t_i)$ for which the different samples $s(t_i)$ are identically and independently distributed (i.i.d.) Gaussian stochasts will yield i.i.d. Gaussian stochasts for any orthonormal basis. In other words, transforming Gaussian white noise from one orthonormal basis to another does not induce structure. The discrete Fourier transform is an example of such a transformation.

Time-frequency representations can be obtained by applying a digital filterbank to the input signal. Contrary to a coordinate transform associated with a base change, such as a digital Fourier transform, a digital filterbank introduces time-frequency correlations. However, the output of the filterbank of a time shifted signal will always be the time shifted output of the original signal. Hence its time-frequency representation is time invariant, which facilitates pattern discovery. In contrast time invariance is lost if we time shift a signal and then map it to an orthonormal basis. 

For continuous wavelet transforms it is possible to reconstruct an original signal from the wavelet analyses~\cite[]{Calderon1964,Grossmann1984,Mohlenkamp2008}, this despite a lack of orthonormality. This shows that, in this first processing step, information can in principle be preserved. In practical applications, however, the number of frequencies used will be limited and only information represented at these frequencies is preserved. An extra complication arises in gammatone and gammachirp filterbanks, where, to more closely mimic cochlea behavior, a range of wavelet representations is used~\cite[]{Reimann2011}. Nevertheless, successful, i.e., perceptually similar, signal resynthesis can still be achieved~\cite[]{Irino1999,Hohmann2002} demonstrating that these filterbanks, as far as perception is concerned, largely preserve the original information contained in the signal.

In conclusion, we will use Gaussian white noise signals to determine the filterbank induced correlation distances in the time-frequency representation. Depending on the time-frequency representation, these can either be calculated analytically or estimated from the time-frequency representation of a long white noise signal. Here we use numerical estimates. The correlation distances in the time and frequency direction define natural units on the time-frequency plane. We propose that sound structure can be inferred if a pattern in the filterbank output extends beyond the correlation distance induced by the filterbank. Put differently, significant --- source physics indicative -- structure is associated with redundancy beyond the white noise correlation distance. This is typically a repeated energy modulation pattern in some arbitrary time-frequency plane direction. In this paper we analyze the time and frequency direction separately. We calculate the extent the log energy at one point in the time-frequency representation deviates from that one correlation distance away in either the temporal or the frequency direction. If similar --- redundant --- deviations are found in neighboring points of the time-frequency representation and if these constitute a 'tract' stretching over several correlation distances, they contribute reliably to the source indicative features we term 'tracts' or 'tract features'. We will refer to an analysis of sound in terms of tract features as the 'texture' of the signal. 

\section{Methods}

In this section we first briefly describe the gammachirp based filters we used to create the time-frequency representation on which our structure extraction is based. Then we introduce 'oriented center surround ratios' ($O_\ov$ and $O_\oh$), which detect energy variations in the frequency and the time direction respectively. In broadband noise, neighboring areas of the time-frequency representation acquire a variety of oriented center surround ratios, while for spectro-temporally localized sound structures we expect alignment of similarly oriented center surround ratios. We therefore average the squared oriented center surround ratios over 'tracts' perpendicular to the direction in which we measured the oriented center surround ratios. With structure present, the squared $O_\ov$'s or $O_\oh$'s add up and produce high amplitude 'tract averages' ($T_-$ and $T_|$), indicative of the presence of tracts of similar $O_\ov$'s or $O_\oh$'s. Thus we expect that high values of $T_-$ and $T_|$ allow us to infer the presence of tones and pulses, respectively.  
  
\subsection{Gammachirp filterbank}
In computational auditory scene analysis (CASA), gammachirp filter banks~\cite[]{Irino2001} are widely used to mimic mammalian cochlear processing. We use a filterbank comprising of 133 segments $s$ (channels), each defined by a gammachirp. A gammachirp is a wavelet with a frequency dependent shape. We use:
\begin{equation}
\gamma_s(t)= N_s t^{n-1} \exp(-2\pi b_1 (Erb_1 f_s+ Erb_0) t ) \exp( i 2 \pi f_s t + i c_1 \ln(t)) \theta(t)
\end{equation}
with $n=4$, $Erb_0=24.7$, $Erb_1=0.0779$, $b_1= 0.707$ and  $c_1=-3.70$, where all settings are adapted from \cite[]{Irino2001,Glasberg1990} except they use $b_1=2.02$ and $Erb_1=0.109$. The frequency dependent normalization $N_s$ is chosen to yield a energy response to a short pulse which is approximately flat in the frequency direction, i.e.
\begin{equation}
\frac{1}{N_s}= f_s \sqrt{\sum_{t_i} \gamma_s(t_i)\gamma_s^*(t_i)}
\end{equation}
Our parameter settings lead to elongated filters, which improve resolution in the frequency direction at the cost of resolution in the temporal direction. In our experience these settings facilitate interpretation of the time-frequency representation. Our 133 frequencies are logarithmically spaced from the minimum frequency $f_{min}=40 Hz$ to the maximum frequency $f_{max}=11025$. That is we have $f_s=f_{min} e^{\alpha (2s-1)/(2n_{seg})}$, where $n_{seg}$ specifies the number of segments used, $\alpha=ln(f_{max}/f_{min})$ and the segments are numbered $s=1,...,n_{seg}$.

The complex amplitude $A(S; t,f)$ of the signal $S$ at location $t,f$ in the time-frequency representation is obtained from $S$ through convolution of $S$ with the gammachirp filter $\gamma_s$: 
\begin{equation}
 A(S; t,f)=  \sum_{t'>0} S(t-t')\gamma_s(t') 
\end{equation}
where $t'$ runs over all the samples between $t=0$ and the time $t_{max}=0.4 s$ corresponding to the maximum filter length. We used a fast Fourier transform overlap-and-add implementation to calculate $A(S; t,f)$. Using $A$ we define the energy of the signal $S$ at location $t,f$ as:
\begin{equation}
 E(S; t,f)=  A(S; t,f) A^*(S; t,f).
\end{equation}
From which the energy in decibels (dB) is calculated as:
\begin{equation}
 E_{dB}(S; t,f)=  10 \log_{10}( E(S; t,f) ).
\end{equation}
In our experience using log energy provides a form of dynamic range compression which yields better results then working directly with energy. The likely explanation for this is the large dynamic range exhibited by natural sounds, which spans many orders of signal intensity. 

A note on notation: We write for example $E(S; t,f)$ to indicate that the representation $E$ has been derived from the representation $S$, while $t,f$ indicate the location in the time-frequency representation.  Occasionally we will drop $t$ and $f$ to indicate usage of the full data in the representation, e.g. $E(S)$ or we drop the explicit reference to the representation from which it was derived and simply write $E(t,f)$. Furthermore, we will use $W$ to denote a white noise signal, and hence $E(W;t,f)$ is the energy of a white noise signal $W$ at location $(t,f)$ in the time-frequency representation. We expect that in general it will be clear from the context which notation we use, and we aim to use a notation that best illustrates the step that we make at a specific point in the paper. 

\subsection{Introducing natural units on the time-frequency plane.}

In the digital signal processing implementation of time-frequency analysis we tend to focus on parameters like sample frequency and frame size in the temporal domain and frequency resolution in the frequency domain. Although important signal processing parameters, they are from a physical point of view only discretization parameters required to bring the time-frequency representation in digital form. Apart from imposing resolution limits they have no significance in relation to source physics. 

As natural units of analysis we propose to use the correlation distances in the time-frequency representation. These are estimated directly from the time-frequency analysis of white noise and act as frequency dependent units of time and frequency. Within a correlation distance we expect predominantly filter induced structure, while outside the correlation distance we expect signal induced structure. If we establish a pattern exceeding the local correlation distance, then we conclude with (measurable) confidence that we have found --- physically relevant --- time-frequency structure in the signal. 

These natural units are frequency dependent but are independent of discretization parameters. Naturally, the discretization should be sufficiently fine grained to allow for accurate estimation of the correlation distances. It is possible to derive numerically exact expressions for correlations in the energy representation of gammachirp functions . We are not aware, however, such methods exist for the log energy. Hence we resort to numerical estimates.

The correlation matrix $R_f$ for a time-frequency representation of the log energy $X$ at a frequency $f$ is given by:
\begin{eqnarray}
R_f(X; \delta t,\delta f)&=&\frac{\mu_f([X-\mu_f(X)][X_{\delta t,\delta f}-\mu_{f+\delta f}(X)])}{\sigma_f(X)\sigma_{f+\delta f}(X)}\nonumber\\
\end{eqnarray}
where $X_{\delta t,\delta f}$ stands for the frequency and time shifted $X$ matrix $X_{\delta t,\delta f}(t,f)=X(t+ \delta t,f+\delta f)$. In practice the finite number of frequencies and frames makes it impossible to shift the full matrix. Therefore, in practical situations  $R_f(X;\delta t,\delta f)$ is only defined if $f+\delta f$ is within the defined range of frequencies and $\delta t$ is small with respect to the total time span included in $X$.  The $\mu_f$ and $\sigma_f$ indicate the mean and variance over time at a frequency $f$, i.e., the frequency coordinate is fixed and we average over all elements with the same frequency but distinct time.

We use $R_f$ to define the correlation distances in the log energy representation $E_{dB}$ as:
\begin{eqnarray}
\epsilon_f(f)&=& \inf_{\delta f \geq 0}: R_{f}(E_{dB}; 0,-\delta f) < \theta    \nonumber\\
\epsilon^f(f)&=& \inf_{\delta f \geq 0}: R_{f}(E_{dB}; 0,\delta f) < \theta     \nonumber\\
\epsilon_t(f)&=& \inf_{\delta t \geq 0}: R_{f}(E_{dB}; -\delta t,0) < \theta  \nonumber\\
\epsilon^t(f)&=& \inf_{\delta t \geq 0}: R_{f}(E_{dB}; \delta t,0) < \theta 
\end{eqnarray}
with $\theta < 1$ a conveniently chosen threshold value, in this paper we use $\theta=0.2$. On the basis of symmetry we expect that $\epsilon_t(f)=\epsilon^t(f)$. Because we have chosen logarithmically spaced frequencies we also expect $\epsilon_f(f)=\epsilon^f(f)$ for a true wavelet representation. However because we prefer to work with biological inspired time-frequency representation, which lacks true frequency invariance, this relationship does not apply and we need to treat them separately. We, therefore, find it more elegant to also treat $\epsilon_t(f)$, $\epsilon^t(f)$ separately. In our code we improve the estimated $\epsilon$'s further by interpolating between the $\epsilon$'s found in the discretized representation and its suprathreshold neighbor. Such a neighbor exists because $R_{f}(X; 0,0)=1$ and hence larger than the threshold. Based on these distances we also introduce the threshold crossing vectors:$\boldsymbol\epsilon_f= (0,-\epsilon_f)$,  ${\boldsymbol\epsilon}^f=(0,\epsilon^f)$, ${\boldsymbol\epsilon}_t=(-\epsilon_t,0)$, and ${\boldsymbol\epsilon}^t=(\epsilon^t,0)$.

Given a time-frequency representation $X$, the threshold $\theta$ is one of only three free parameters of the structure extraction algorithms we propose; $\theta$ determines the separation between center and surround of the oriented center surround ratios. 

\subsection{Oriented center surround ratios}

Any repeated pattern in the log energy occurring well above chance indicates source imposed structure. This paper focuses on establishing simple local structures like pulses and tones. Pulses are visible in the cochleagram as vertical ridges of higher energy, while tones are visible as horizontal ridges. The log energy values on a ridge are higher than on the ridge's surroundings. This suggests the notion of oriented center surround ratios;  these measure the variation perpendicular to a ridge and yield high values on top of the ridge. We define, in either the temporal or the frequency direction, the average surrounding as the average log energy of two time-frequency points a single correlation distance away from the time-frequency point under study. In mathematical form we thus obtain the oriented center surround ratios:
\begin{eqnarray}
O_\oh(X(S);t,f)= X(S;t,f)-\frac{1}{2}\left\{ X(S;t- \epsilon_t,f)+ X(S;t+\epsilon^t,f)\right\},\nonumber\\
O_\ov(X(S);t,f)= X(S;t,f)-\frac{1}{2}\left\{ X(S;t,f- \epsilon_f)+ X(S;t,f+ \epsilon^f)\right\}.
\end{eqnarray}
for the temporal and frequency direction respectively. For notational clarity we dropped the explicit reference to $f$ in the correlation distances $\epsilon$ because the frequency is clear from the context, furthermore we wrote $X(S)$ to denote the matrix with elements $X(S;t,f)$. We refer to the result as ratios because subtraction in the log energy domain corresponds to taking ratios in the energy domain. This is analogous to the use of 'ratio' when expressing the 'signal-to-noise ratio' in decibels.
 
\subsection{Temporal and frequency tracts}
Often the isolines of oriented center surround ratios in the time-frequency plane co-occur  in bundles --- 'tracts' --- of similar orientation. Although values along these lines are constant, the values in the perpendicular direction can vary considerably and can change sign within a correlation distance. This is because the oriented center surround ratios are proportional to the second directional derivatives of the log energy. Suppose for  example that a tone is present. If we inspect the 'vertical' development of $O_\ov$ while crossing a horizontal ridge, we first see it become negative, then positive, and then negative again. If, therefore, we simply average $O_\ov$, contributions of opposite sign cancel each other although they are indicative of a similar orientation. To avoid these cancellations we compute the root-mean-square of $O_\oh$ and $O_\ov$, both in the direction perpendicular and parallel to their orientation. 

\begin{figure}[!ht]
\begin{center}
\setlength{\unitlength}{1.0cm}
\begin{picture}(5,5)
\put(0.0,0.0){\includegraphics{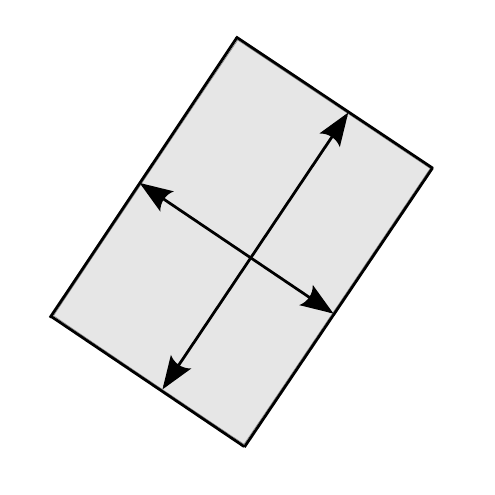}}
\put(2.8,2.35){{$(t,f)$}}
\put(3.5,4.0){{$\boldsymbol{w_2}$}}
\put(0.7,3.3){{$\boldsymbol{v_1}$}}
\put(1.0,0.7){{$\boldsymbol{w_1}$}}
\put(3.5,1.7){{$\boldsymbol{v_2}$}}
\put(4.5,0.8){\vector(0,1){0.5}}
\put(4.5,0.8){\vector(1,0){0.5}}
\put(4.2,0.9){{$f$}}
\put(4.6,0.45){{$t$}}
\end{picture}
\end{center}
\caption{Parametrization for time-frequency boxes  $\Diamond(\boldsymbol{v_1},\boldsymbol{v_2};\boldsymbol{w_1},\boldsymbol{w_2};t,f)$ used for pattern tract detection. For the calculation of $T_{|}$ we use $\boldsymbol{v_1}=c_p\boldsymbol{\epsilon_t}$, $\boldsymbol{v_2}=c_p\boldsymbol{\epsilon^t}$, $\boldsymbol{w_1}=c_t\boldsymbol{\epsilon_f}$,and $\boldsymbol{w_2}=c_t\boldsymbol{\epsilon^f}$. For the calculation of $T_{-}$ we use $\boldsymbol{v_1}=c_p\boldsymbol{\epsilon_f}$, $\boldsymbol{v_2}=c_p\boldsymbol{\epsilon^f}$, $\boldsymbol{w_1}=c_t\boldsymbol{\epsilon_f}$, and $\boldsymbol{w_2}=c_t\boldsymbol{\epsilon^t}$. 
With these choices these time-frequency boxes $\Diamond$ for a given frequency $f$ are completely defined by the scalar parameters $c_p$, $c_t$ and the frequency dependent threshold crossing vectors $\boldsymbol{\epsilon_f}$, $\boldsymbol{\epsilon^f}$, $\boldsymbol{\epsilon_t}$, and $\boldsymbol{\epsilon^t}$. The latter are fully determined by the choice of the threshold $\theta$.}
\label{fig:BBox}
\end{figure}
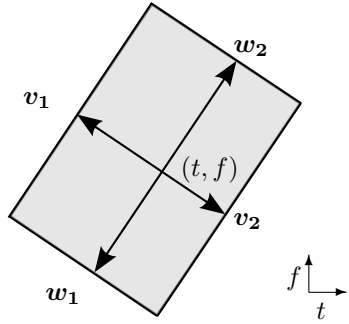

As addressed earlier, to find structure beyond that introduced by the filters we need to average over a distance exceeding the correlation distance in the --- tract --- direction perpendicular to the oriented center surround ratios (denoted by $c_t$), while in the parallel direction we average over a distance smaller or similar to the correlation distance (denoted by $c_p$). Averaging in the perpendicular direction integrates evidence for oriented structures in the time-frequency representation, while  parallel averaging ensures that tracts which are only approximately perpendicular still contribute. This leads to our squared tract features:

\begin{eqnarray}
T_{|}^2(O_\oh;t,f)= \mu_{\Diamond(c_p\boldsymbol{\epsilon_t},c_p\boldsymbol{\epsilon^t};c_t\boldsymbol{\epsilon_f},c_t\boldsymbol{\epsilon^f};t,f)}(O_\oh^{\stackrel{2}{.}})\nonumber\\
T_{-}^2(O_\ov;t,f)= 
\mu_{\Diamond(c_p\boldsymbol{\epsilon_f},c_p\boldsymbol{\epsilon^f};c_t\boldsymbol{\epsilon_t},c_t\boldsymbol{\epsilon^t};t,f)}(O_\ov^{\stackrel{2}{.}}).
\end{eqnarray}
where the dot under the exponent indicates that we take the exponent element-wise and not matrix-wise, the $\Diamond$ indicates the area used for taking the mean $\mu$ (see figure~\ref{fig:BBox}). An area $\Diamond$ depends on the four vectors connecting the current time-frequency $(t,f)$ to the correlation threshold crossing point in the temporal and the frequency direction, and it also depends on two parameters that control the size of this averaging area in the pattern $c_p$ and the tract direction $c_t$.  Each combination of a scaled pattern threshold vector (in the case of $T_{|}$ these are given by $c_p \boldsymbol{\epsilon_t}$ or $c_p \boldsymbol{\epsilon^t}$) with a scaled tract direction threshold vector (in the case of $T_{|}$ these are given by $c_t \boldsymbol{\epsilon_f}$ or $c_t \boldsymbol{\epsilon^f}$) defines a parallelogram which is part of the area $\Diamond$. 

For the tract calculation in this paper we use $c_p=0.7$ (smaller than a correlation distance in the pattern direction) and $c_t=2$ (greater than a correlation distance in the tract direction). We have only three free parameters --- $\theta$, $c_p$, and $c_t$ --- in the presented structure extraction algorithms. Interestingly, all pertain to the how we use and define the noise correlation distance. 

The extraction of oriented center surround ratios and tracts from the log energy representation is performed by dedicated \verb!C++! code. For ease of use, this code is wrapped in Python.

The underlying code is made available under the Apache License, Version 2.0 as an open source repository through 
GitHub: \url{https://github.com/soundappraisal/libsoundannotator}. Documented code examples are available through GitHub licensed under 
the same Apache License, Version 2.0: \url{https://github.com/soundappraisal/soundannotatordemo}.

\section{Experiments and Results}

In this section we address two experiments. In the first experiment we show that our features can identify tonal, pulsal, and (broadband) noisy sonic textures in the time-frequency plane. For this we calculate the tract features on a long (Gaussian) white noise signal -- our definition of an unstructured signal -- and we calculate, visualize, and analyze the tract features of three isolated environmental sounds used in the perceptual experiment by \cite{Gygi_2007}. By comparing the unstructured reference sound with the (more structured) environmental sounds we can identify time-frequency regions that differ statistically from Gaussian noise and that --- in our definition --- are therefore structured. We picked 'clapping hands', 'a siren', and 'rain'. These are each a representative example of the three different perceptual clusters apparent from the perceptual experiment that Gygi referred to as 'discrete impact', 'harmonic', and 'continuous' sounds, respectively; a classification similar to what we propose. 

In our second experiment we compare the separation in sonic textures corresponding to noise, pulses, and tones for Gygi's full database and we present the resulting classification in discrete impact, harmonic, and continuous sounds. In his experiment \cite{Gygi_2007} had asked his participants to rate the similarity of 100 short and isolated environmental sounds, which led to a perceptual similarity matrix. With multidimensional scaling he identified the main perceptual dimensions of which the main two allowed him to to separate the resulting (perceptual) plane into regions for discrete impact, harmonic, and continuous sounds. We will achieve similar results but in our case with "objectively" estimated pulsality, tonality, and noisiness descriptors based on our tract features.  

We can not do this directly because our tract features are defined for each point of the time-frequency plane, while Gygi's perceptual dimension are defined on sound file level. Therefore we use the tract features to define three sound file level 'descriptors' that reflect a sound file's 'pulsality', 'tonality', and 'noisiness'. The definition of these descriptors requires us to pick thresholds. Inspection of the data from the first experiment suggested suitable threshold values. We investigate two variants of these sound file descriptors. In the first variant the tract features are used to attribute cochleagram energy to pulsality, tonality, and noise. In the second variant energy is ignored and the tract features are used to assign cochleagram areas to pulsality, tonality, noisiness. Once defined, the features for all environmental sounds are calculated, and we correlate the resulting pattern with the perceptual MDS dimensions found by \cite{Gygi_2007}.

\subsection{Experiment 1: White noise and environmental sounds tract features}
  
In this first experiment we calculate the tract features for three environmental sounds and for white noise for comparison. Due to their construction, the environmental sounds used in the perceptual experiment by \cite{Gygi_2007} have prolonged periods in which the samples take are equal to $0$, this leads to zero (local) energy which implies that our features, involving log energy, are not well defined in these areas. To circumvent this problem we added a white noise 'floor' to the environmental sounds drawn from the same distribution as we will use for creating our white noise recording before processing (see the next paragraph). Before processing the cochleagram, we further reduced the number of data points by only including each 50st sample in the energy representation. Dubious as this might be from a signal processing perspective, we hardly see an impact of this additional decimation on our results. The resulting representation, which also includes a regular downsampling by a factor $2$, has a sampling frequency of $441$ Hz. This corresponds to a cochleagram sample period of about $2.2$ ms, which is still short compared to the typical frame rate of $100$ Hz ($10$ ms) in typical speech recognition systems. After calculating our features we used the Matplotlib module for Python 2.7 to plot the results. 
 
To create reference white noise distributions we processed $200$ seconds of white noise. The white noise is generated by drawing samples from a binomial distribution $B(n, p)$ with $n=20$ and $p=0.5$. The distribution is generated using the Mersenne Twister built into the Numpy module in Python 2.7. After drawing a sample we subtract $10$ to end up with a value drawn from a distribution centered at $0$. This reference signal was saved in 16 bit signed integer format to a wav-file using the wave module from Python's standard library. This wav-file was then processed with the same code used to also process the environmental sounds. 
                                                                                                                                                                                                                                                                                                                                           
\subsection{Result 1: White noise and environmental sounds tract features}

\begin{figure}[!ht]
\begin{center}
\includegraphics[width=0.85\textwidth]{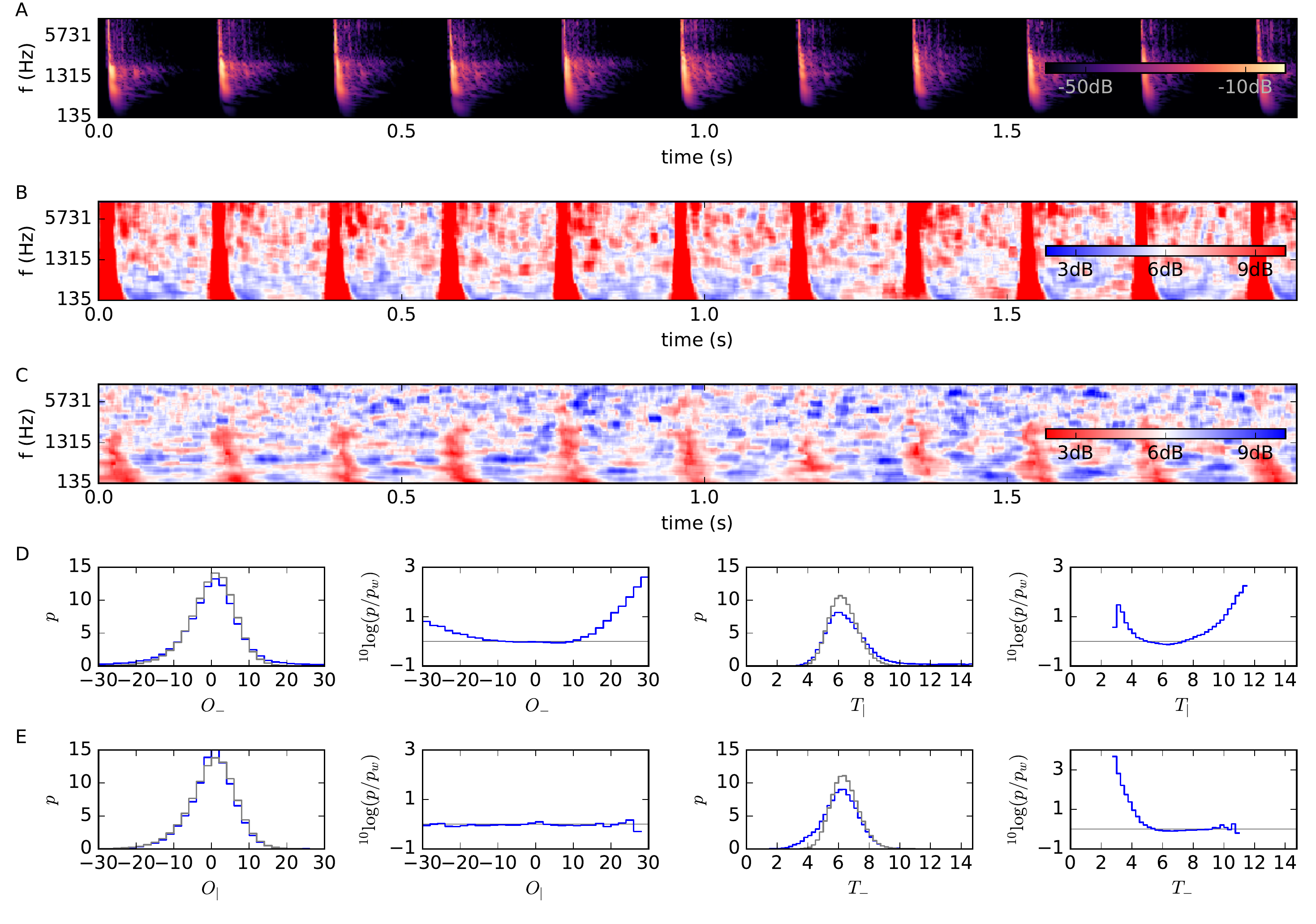}
\end{center}
\caption{Cochleagram, Oriented center surround ratios and tract feature distributions calculated on a recording of a person clapping hands, the recording was selected from the perceptual study as an example of an impact sound. A. Cochleogram of the sound showing the energy distribution (in dB relative to the maximum) over frequency and time, B. vertical tract feature (in dB), C. horizontal tract feature (in dB), D. comparison of vertical tract feature related distributions calculated on white noise (light gray) and on the sound (blue) from left to right: normalized histogram of horizontal center surround ratio, log prevalence ratio of these sound and white noise histograms, normalized histogram of vertical tract feature values, log prevalence ratio of these sound and white noise histograms; E. comparison of horizontal tract feature related distributions calculated on white noise (light gray) and on the sound (blue) from left to right: normalized histogram of vertical center surround ratio, log prevalence ratio of these sound and white noise histograms, normalized histogram of horizontal tract feature values, log prevalence ratio of these sound and white noise histograms.}
\label{fig:clapsa}
\end{figure}

\begin{figure}[!ht]
\begin{center}
\includegraphics[width=0.85\textwidth]{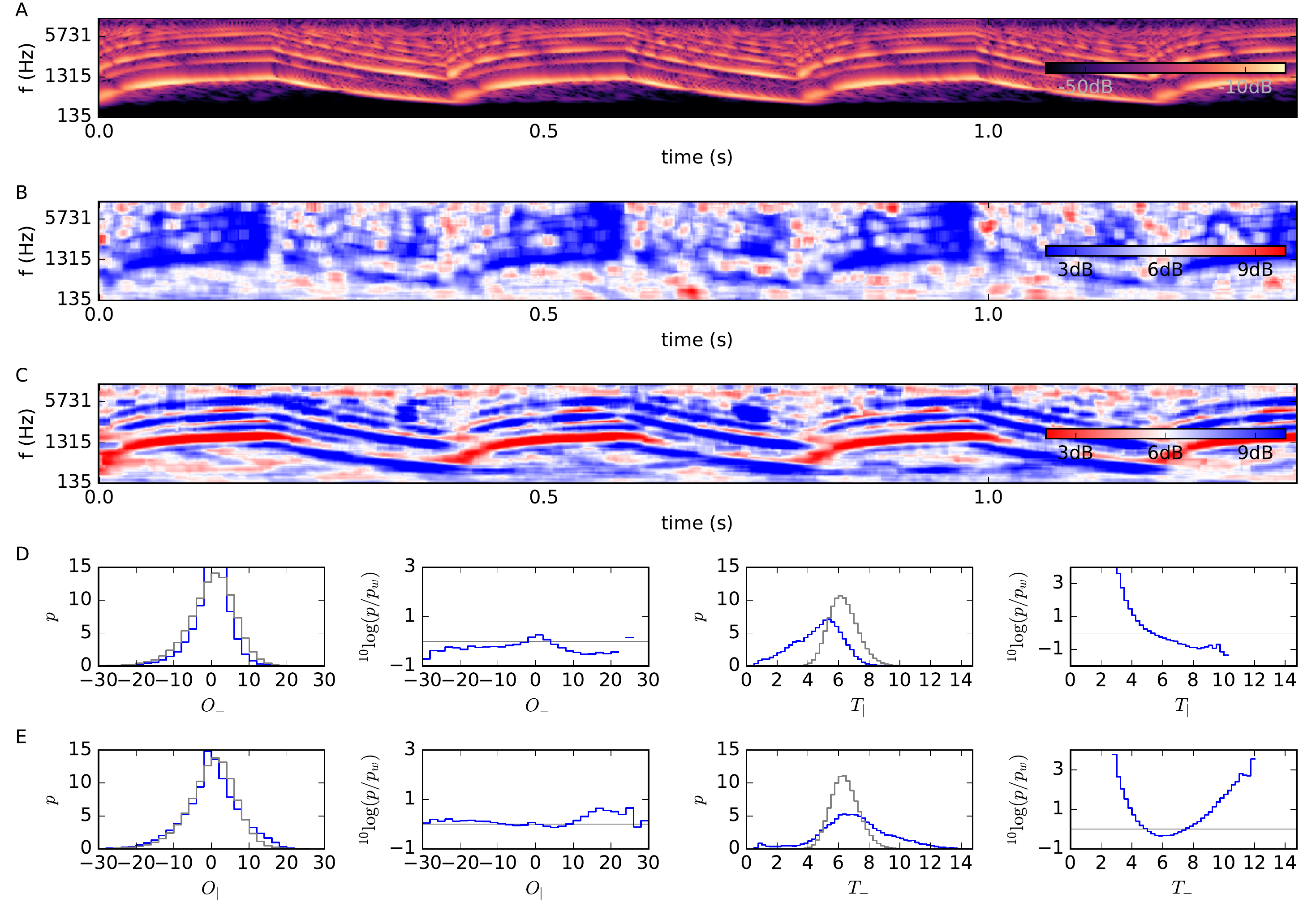}
\end{center}
\caption{cochleagram, Oriented center surround ratios and tract feature distributions calculated on a recording of a siren, the recording was selected from the perceptual study as an example of a harmonic sound. Subpanels are as in~figure(\ref{fig:clapsa}) }
\label{fig:sirene}
\end{figure}

\begin{figure}[!ht]
\begin{center}
\includegraphics[width=0.85\textwidth]{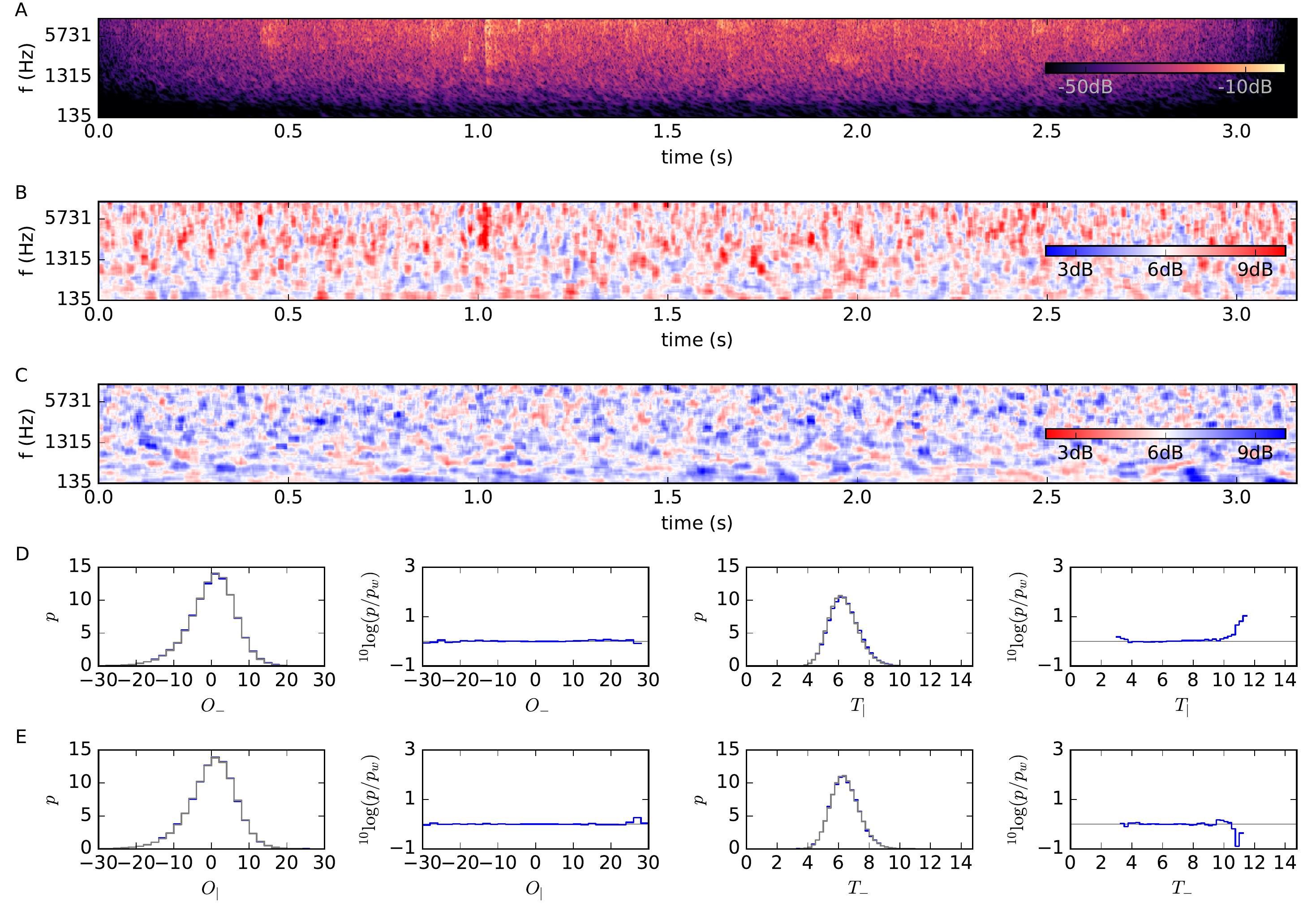}
\end{center}
\caption{cochleagram, Oriented center surround ratios and tract feature distributions calculated on a recording of a rain, the recording was selected from the perceptual study as an example of a continuous sound. Subpanels are as in~figure(\ref{fig:clapsa}) }
\label{fig:raina}
\end{figure}

The outcomes of our tract feature calculation are shown in the figures~\ref{fig:clapsa}, figure~\ref{fig:sirene}, figure~\ref{fig:raina}. Figure~\ref{fig:clapsa} shows at the top a cochleagram and the tract features calculated from a recording of hand clapping, this is an example of a sound recording categorized by \cite{Gygi_2007} as discrete impact sounds.  

We use different color coding schemes for the three top panels. Each is visible as a small inset at the right. For the cochleagram (top image) we use the magma colormap, which develops from black to yellow so that similar relative energy differences in dB match similar visual differences~\cite[]{Walt2015, Machado2009, Luo2006}. The second and third panel from the top reflect the prevalence of vertical (pulsal) and horizontal (tonal) tract features respectively. The color map (red-white-blue) for these panels has been chosen so that red is indicative of pulses and counter indicative of tones, while blue is counter indicative of tones and indicative of pulses. This entails that the colormap for the pulsal tract features has high values (red) for pulses, while the colormap for the tonal tract features is swapped to that high values (blue) indicate tones. In both cases the middle of the scale corresponds to the absence of evidence of tones and pulses, which is indicative for noisy contributions.

The cochleagram in the top panel of figure~\ref{fig:clapsa} A shows each clap as a light vertical streak followed by (noisy) reverberations that die out at a constant rate. Most of the time-frequency plane reflects little energy and is dark. 

The pulsal tract feature $T_|$ in panel B shows, as intended, very clear red, pulsal, patches corresponding to each clap. At high frequencies, where individual contributions can be identified with high temporal resolution, pulsal evidence after each clap indicates its first resolved echoes. Most of the clap's reverberation is noisy (due to the confluence of many time-delayed echoes of the clap) and is apparent as a "noisy" fine-structure. Noisy fine-structure dominates everywhere the signal is neither tonal nor pulsal. 

The tonal tract feature $T_{-}$ is shown in panel B. In this case blue indicates high tonal information. Although less pronounced, the individual claps are visible as red low tonal structures in the low frequency range. This makes sense because the local dominance of the pulse is counter evidence for tones.  We also see some horizontal red and blue streaks in the tonal tract features plot, as these can not be attributed to tones in the input signal they have an other origin. In the presence of resonances in the room acoustics we expect reverberation at these resonances to be more energetic than at non-resonant frequencies, thus forming horizontal ridges and leading to an enhanced tonal tract feature. 

In the lower part of the figure we present two rows of four subfigures that summarize pulsal and tonal evidence of the whole signal. The first row, marked D, presents information about vertical tract features, while the lowest row, marked E, presents information about the horizontal tract features. The four subfigures on the left present center surround ratios ranging from -30 to 30 dB. The four subfigures on the right present the tract feature information. Since this is the root-mean-square of the center surround ratios it has an all positive range. The first and third column shows the histogram of the center surround ratios and the tract features of the clap signal in blue and, for reference, the corresponding distribution of a 200 s white noise signal in gray. 

The center surround ratios histograms in the first column step from -30 dB to 30 dB in steps of 2 dB. The area under each histogram sums to 100, so that the $p$ on the vertical axes reflects probability as percentage. The tract feature histograms in the third column step from 0 to 15 in steps of 0.25 dB, again so that the area under the histograms sums to 100 and $p$ reflects a percentage. The tract features show a peak close to 6 dB. 

The second and fourth column present the base-10 log-ratio of the sound and the white noise reference. This corresponds to prevalence ratios of the oriented center surround and tract feature values falling in to the same bin for white noise and the hand clapping recording. The prevalence ratios gives an impression of the probability that the distribution of values occurs in white noise. 

Assuming the 200 s of white noise give a sufficiently accurate description of the white noise distribution, we can conclude from this figure that values of $O_\ov$ between 28 and 30 occur $10^{2.6}\approx 400$ more often in the vertically oriented center surround ratio for the hand clapping recording than expected if the recording represented white noise. Because neighboring point in the time-frequency plane are correlated we can not calculate the corresponding probability ratios with simple probability arguments, however these correlations are also present in the white noise reference, which makes it in our view unlikely that the large differences observed can be attributed to mere chance. Therefore the presented values represent a reasonable indicator of the actual probability of this center surround ratio. 

In figure~\ref{fig:clapsa}.E on the right we see that prevalence ratio for the horizontal tract feature calculated on the siren recording reach a value close to $10^4$. The graph stops due to the absence of lower values in the white noise reference data. The scarcity of these values also explains why the graphs are less smooth at these low values: the error in the estimation of the white noise prevalence translates into errors in the estimated ratios. These errors can be estimated by rerunning the white noise analysis with several different realizations of the white noise, however our aim here is not accurate estimates of the prevalence ratios, but to demonstrate that the tract features contain information similar to those used by humans. 

Somewhat ironically here these low values --- and not the high values we were aiming for --- form an illustration of the ideas behind our transition from oriented center surround ratios to tract features. The vertical center surround feature shows a slightly enhanced prevalence of values close to $0$ as compared to the prevalence in white noise. However the prevalence ratio is still close to $1$, so observing a value close to $0$ doesn't reliably indicate the presence of a pulse. However after applying the averaging over a time-frequency box resulting in the horizontal tract feature, the horizontally aligned low values give rise to low horizontal tract features and the prevalence ratio at these low values reaches over a $1000$.

The organization of figure~\ref{fig:sirene} is analogous to the organization of figure~\ref{fig:clapsa}, but the sound analyzed is a siren, which belongs to the group of harmonic sounds in the classification of Gygi and coauthors. The siren is a frequency modulated signal. The vertical tract feature takes on low values, smaller than 5, for a large part of this representation, thus signaling the absence of pulses and the presence of tones. This was marginally visible in the horizontal oriented center surround ratio, which showed prevalence ratios under 1.5. But we find for this sound prevalence ratios up to $10^4$ for the low vertical tract feature values. 

If we inspect the horizontal tract feature we conclude that it has both large areas with high values and large areas with low values. This might be surprising at first sight as one might expect values to be high for both situations. The origin of this effect is the slower response of the gammachirp filters at lower frequencies. In accordance with biology the filters are causal and thus the low frequency filters have their maximum response later than might be expected when only taking the signal into account. Of course one can use a filterbank in which maximum responses are aligned to get rid of this effect, but this entails introducing non-causality. 

In the vertical oriented center surround values we see prevalence ratios slightly larger than 1 at large deviations from $0$ and around $0$ aswell. At intermediate values the ratios are somewhat smaller smaller than $1$. In the transition from the vertically oriented surround feature to the horizontal tract feature, we see that prevalence ratios are boosted at high and low values are boosted to values over a $1000$. Again illustrating that the tract features have the ability to accumulate evidence for the presence and absence of tones and pulses. 

The organization of figure~\ref{fig:raina} is again analogous to the organization of figure~\ref{fig:clapsa}. This figure analyses recorded rain noise, which Gygi and coauthors classified as a continuous process sound. Rain is a stochastic process and if the individual droplets can no longer be resolved the drop sounds constitute a broadband noise. If we compare the vertical and horizontal oriented center surround ratios we only see minimal differences between the rain and the white noise histograms. Accordingly the prevalence ratios for these features are all close to $1$. 

Inspecting the tract features however, we see prevalence ratios up to about 10 for the vertical tract feature and prevalence ratios smaller than 1. In both cases at values higher than 10 of the tract feature. For lower values the prevalence ratio is close to 1. It can be suspected that these deviating values are an artefact due to the low number of observations involved. However, if we inspect the vertical center surround feature in the second panel from the top we observe that high (red) values are more frequent at high frequencies. This is an indication that individual droplets can still be partly resolved at higher frequencies.

All in all we can conclude that the center surround ratio and the tract features behave as expected and by and large according to the design requirements. In addition it seems that the features are sensitive to subtle signal details such as individual clap echoes, room resonances, and the detailed statistics of noise resulting from numerous droplets.

\subsection{Experiment 2: Comparison tract feature with perceptual data. }

A direct comparison of the tract features with human physiological or perceptual data is, to our knowledge, still impossible. We have therefore chosen to construct single-valued tract-based sound descriptors for each recording in Gygi's database and to use these values to characterize and compare sounds. We expect that our characterization conforms with the categorization in harmonic (tonal), impact (pulsal), and continuous process (broadband noise) sounds  \cite{Gygi_2007}. We implemented this by assigning either the energy or the surface area of time-frequency plane regions to pulsality for high vertical tract feature values, to tonality for high horizontal tract feature values, and to noise if both values are low. The resulting values are normalized either by dividing by the total energy or by dividing by the total included area. Because humans are more sensitive to relative energy differences rather than absolute differences, we apply a logarithm to model Weber's law (see for example: ~\cite{kandel1985principles}). For each file this leads to a measure of pulsality (P), tonality (T), and noisiness (N) that we formally define as:
\begin{eqnarray}
P(E) &= & \log\left(\sum_{t,f}\sigma(T_{|}\ -\Theta) E\right)-\log\left(\sum_{t,f} E\right)\nonumber \\
T(E) &= & \log\left(\sum_{t,f}\sigma(T_{-}\ -\Theta) E\right)-\log\left(\sum_{t,f} E\right)\nonumber \\
N(E) &= & \log\left(\sum_{t,f}\left(1-\sigma(T_{|}\ -\Theta)\right)\left(1- \sigma(T_{-}\ -\Theta)\right) E\right)-\log\left(\sum_{t,f} E\right)
\end{eqnarray}
Where  $E$ stands for energy if we calculate how the energy is distributed over pulses, tones and nosies. If we replace $E$ with a matrix containing $1$ at each time-frequency location, we calculate the area occupied by pulses, tones and broadband noises in the time frequency plane. We will present results for both these cases.

We used the following sigmoid function $\sigma$:
\begin{equation}
 \sigma(x)=(1+\tanh( 2 s x))/2
\end{equation}
where we denote with  $s$ the slope of the sigmoid at  at the origin $(x=0)$ and we have set $s=2.5$. We picked the thresholds through visual inspection of the histograms in experiment 1. We originally estimated that values of $T_{|}$ and $T_{-}$ higher than $\Theta=8$ are indicative of structure, this is roughly the point were we observe a transition from white noise dominated to structure dominated in the histograms. For this study we performed no optimization of this parameter. Looking at the 
prevalence ratios we can argue that a slightly higher value might be better because we shouldn't look at the point where the histograms cross, but at he point at which prevalence ratios cross the value $2$. This than again might be an under estimate because we don't know the effects of correlations on the enhancement of these prevalence ratios. Alternatively, we could say that there is a probability of about $10 \%$ to find values exceeding $8$ in white noise, and therefore rejecting that these values are white noise corresponds with p-value close to $0.1$.  For the future the threshold $\Theta$ and the slope $s$ are free parameter than can be optimized on available training data. Combined with the three free parameters in the definition of the tract features, and assuming we use the same values for horizontal and vertical tract features, this brings the total number of free parameters to five for this experiment. 

\subsection{Result 2: Comparison tract feature with perceptual data. }

\begin{figure}[!ht]
\begin{center}
\includegraphics[width=\textwidth]{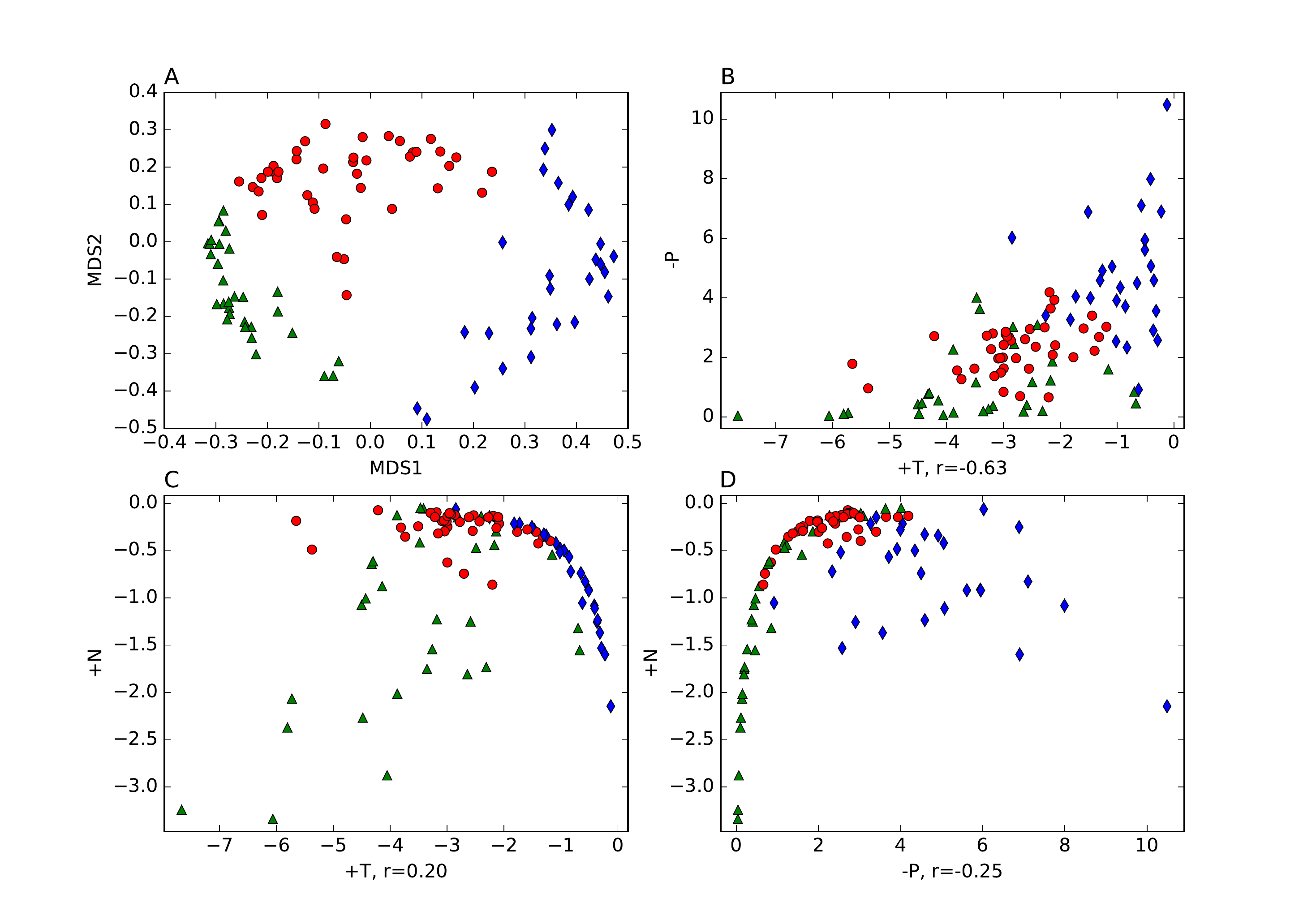}
\end{center}
\caption{Scatter plots of the environmental sounds on the basis of human perceptual space~\cite[]{Gygi_2007} and the energy based pulsality $P(E)$, tonality $T(E)$, and noisiness $N(E)$ . The red circles refer to sounds classified as continuous process sounds, blue diamonds correspond to harmonic sounds, and green triangles correspond to impact sounds. Subfigure A. shows the environmental sounds in the first two MDS dimensions of the human perceptual space. Subfigure B. shows these same sounds in descriptor space spanned by negative pulsality (-P) and tonality(+T). The reversed pulsality on the y-axis (indicated by the minus sign) facilitates comparison with A. Subfigure C. is like B. but based on noise (+N) and tonality (+T) and subfigure D. is based on noise (+N) and reversed pulsality (-P).}
\label{fig:logsumE_off_FS}
\end{figure}

\begin{figure}[ht]
\begin{center}
\includegraphics[width=\textwidth]{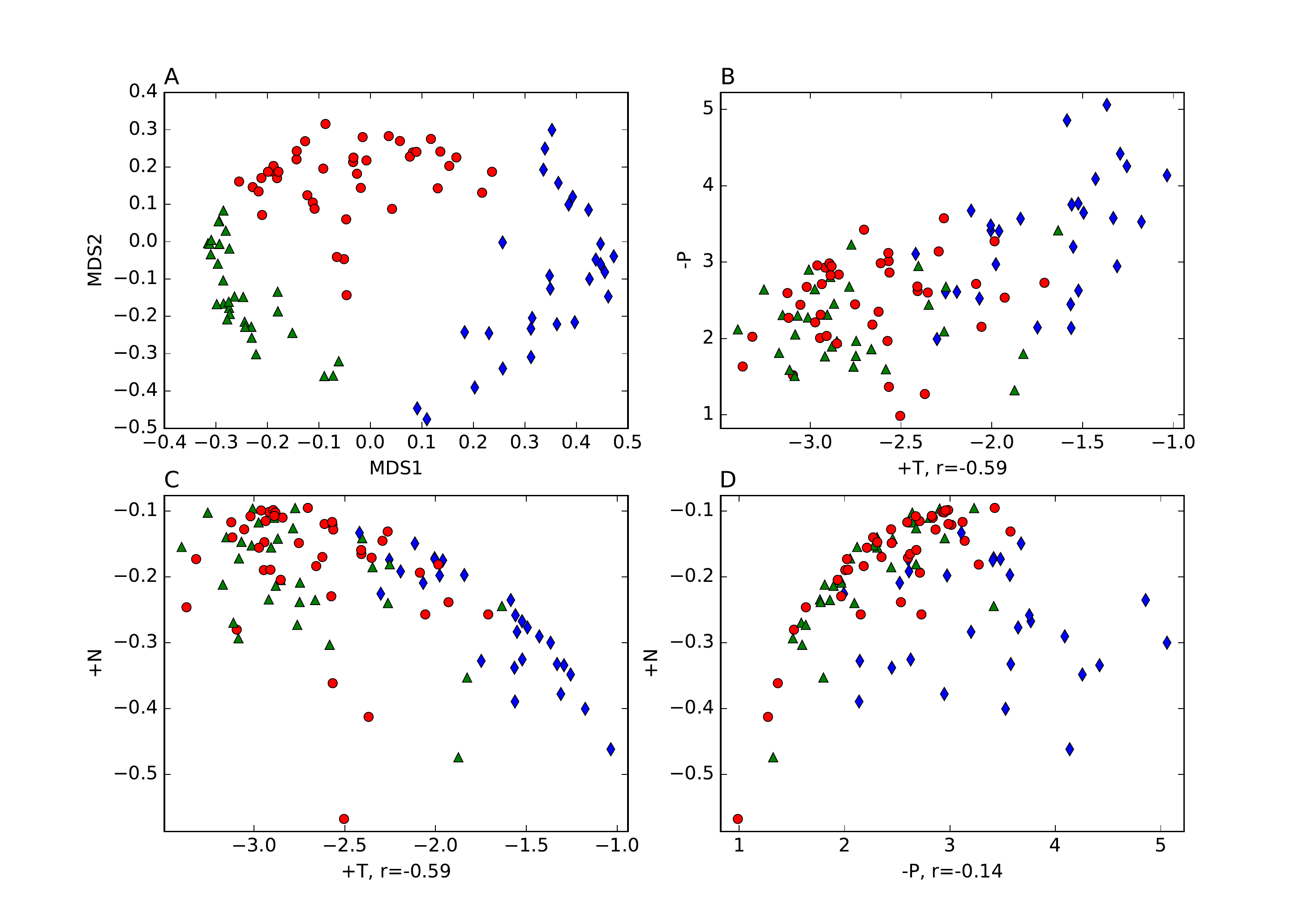}
\end{center}
\caption{ Scatter plots of the environmental sounds on the basis of human perceptual space~\cite[]{Gygi_2007} and the area based pulsality $P(1)$, tonality $T(1)$, and noisiness $N(1)$ . The red circles refer to sounds classified as continuous process sounds, blue diamonds harmonic sounds, and green triangles impact sounds. Subfigure A. shows the environmental sounds in the first two MDS dimensions of the human perceptual space. Subfigure B. shows these same sounds in descriptor space spanned by negative pulsality (-P) and tonality(+T). The reversed pulsality on the y-axis (indicated by the minus sign) facilitates comparison with A. Subfigure C. is like B. but based on noise (+N) and tonality (+T) and subfigure D. is based on noise (+N) and reversed pulsality (-P).}
\label{fig:surface_off_FS}
\end{figure}

The recording descriptors $P(E), T(E), N(E)$ and $P(1), T(1), N(1)$ -- the energy and area based pulsality, tonality, and noisiness -- are used to create scatter plots that position the environmental sounds in 2 D so that they can be compared visually with the perceptual experiment. This information is presented in figure~\ref{fig:logsumE_off_FS} and figure~\ref{fig:surface_off_FS}, respectively. For comparison we also present in subplot A the data on the basis of the first two MDS dimension of Gygi's perceptual experiment. 

We calculated the Pearson correlation coefficients of all recording descriptors with the first three MDS dimensions computed from the perceptual similarity experiment and present these in table~\ref{tab:corrcoeffsenergybased} for energy based and table~\ref{tab:corrcoeffs} for area based pulsality, tonality, and noisiness. These tables also list the Pearson correlation coefficients for recording descriptors calculated previously~\cite{Gygi_2007}.

\begin{table}[!ht]
{\begin{tabular}{ll|ll|ll}\toprule
    \multicolumn{2}{c|} {MDS1}   &\multicolumn{2}{c|} {MDS2}& \multicolumn{2}{c} {MDS3}\\
    Variable& $\left|r\right|$ & Variable &  $\left|r\right|$ & Variable &  $\left|r\right|$ \\
    \midrule
    Mean pitch salience        & 0.75           & RMS band fc=250 Hz        & 0.45          & Duration                    & \textit{0.37} \\
    \textbf{Pulsality, $P(E)$} & \textit{0.74}  & Autocorr. max             & \textit{0.44} & Spectral centroid           & \textit{0.28} \\ 
    \textbf{Tonality $T(E)$}   & 0.71           & \textbf{Noise $N(E)$}     & 0.43          & RMS in band fc 5 8 kHz      & \textit{0.27} \\ 
    Spectrum SD                & \textit{0.61}  & Mean pitch salience       & \textit{0.37} & Adjusted RMS/total RMS      & \textit{0.27} \\ 
    Max pitch salience         & 0.58           & Burst ratio               & \textit{0.34} & RMS in band fc 5 4 kHz      & \textit{0.26} \\
    Modulation spectrum max    & \textit{0.57}  & Median pitch              & \textit{0.33} & RMS in band fc 5 500 Hz     & 0.26          \\
    Spectral skew              & 0.52           & RMS in band fc=500 Hz     & 0.32          & \textbf{Noise  $N(E)$}      & \textit{0.26} \\
    Mean autocorr. peak        & \textit{0.49}  & \textbf{Tonality  $T(E)$} & \textit{0.11} & \textbf{Tonality $T(E)$}    & 0.08          \\
    \textbf{Noise $N(E)$}      & 0.14           & \textbf{Pulsality $P(E)$} & 0.03          & \textbf{Pulsality $P(E)$}   & 0.01          \\ \bottomrule
\end{tabular}}
\caption{Pearson product-moment correlation coefficients  of pulsality $P(E)$, tonality $T(E)$, and noisiness $N(E)$  and competing descriptors with MDS dimensions. The r-values for the competing descriptors are taken from \cite{Gygi_2007}, precise definition of these descriptors or pointers to them can be found there. Correlations printed in italics are negative.}
\label{tab:corrcoeffsenergybased} 
\end{table}

To contextualize these numbers we report the Pearson correlation coefficients between the descriptors:
$P(E)$ and $T(E)$ have $r = -0.63$,  $P(E)$ and $N(E)$ have $r = -0.25$,  $T(E)$ and $N(E)$ have $r= +0.20$, $P(1)$ and $T(1)$ have $r = -0.59$,  $P(1)$ and $N(1)$ have $r = -0.14$,  $T(1)$ and $N(1)$ have $r= -0.60$. These descriptors were constructed to be nearly complementary and they are normalized. Thus we can expect negative correlations: as one descriptor correlates more positively, it forces the others down. And indeed most correlations found are negative. However the correlation between  $T(E)$ and $N(E)$ is slightly positive, this might be due to the tonal tract features becoming high in parts of the timefrequency-plane dominated by reverberation as we saw for the analysis of clapping hands in figure~\ref{fig:clapsa}. Calculation of $p$-values from these and other correlation coefficients presented in this paper is beyond the scope of this paper as we have no theory to identify the true underlying distributions. We do however have visual indications from our scatter plots~(figures: \ref{fig:logsumE_off_FS},\ref{fig:surface_off_FS}) that our $P$,$T$ and $N$ descriptors and the perceptual dimensions don't show normal distributions.

\begin{table}[!ht]
{\begin{tabular}{ll|ll|ll}\toprule
    \multicolumn{2}{c|} {MDS1}   &\multicolumn{2}{c|} {MDS2}& \multicolumn{2}{c} {MDS3}\\
    Variable& $\left|r\right|$ & Variable &  $\left|r\right|$ & Variable &  $\left|r\right|$ \\
    \midrule
    \textbf{Tonality $T(1)$}  & 0.75          & RMS band fc=250 Hz       & 0.45          & Duration                 & \textit{0.37} \\
    Mean pitch salience       & 0.75          & Autocorr. max            & \textit{0.44} & Spectral centroid        & \textit{0.28} \\ 
    Spectrum SD               & \textit{0.61} & Mean pitch salience      & \textit{0.37} & RMS in band fc 5 8 kHz   & \textit{0.27} \\ 
    Max pitch salience        & 0.58          & Burst ratio              & \textit{0.34} & Adjusted RMS/total RMS   & \textit{0.27} \\ 
    Modulation spectrum max   & \textit{0.57} & Median pitch             & \textit{0.33} & RMS in band fc 5 4 kHz   & \textit{0.26} \\
    \textbf{Pulsality $P(1)$} & \textit{0.55} & \textbf{Noisiness $N(1)$}& 0.33          & RMS in band fc 5 500 Hz  & 0.26          \\
    Spectral skew             & 0.52          & RMS in band fc=500 Hz    & 0.32          & \textbf{Tonality $T(1)$} & 0.14          \\
    Mean autocorr. peak       & \textit{0.49} & \textbf{Tonality}        & \textit{0.26} & \textbf{Noisiness $N(1)$}& \textit{0.03} \\
    \textbf{Noisiness $N(1)$} & \textit{0.34} & \textbf{Pulsality $P(1)$}& 0.05          & \textbf{Pulsality $P(1)$}& \textit{0.01} \\
    \bottomrule
\end{tabular}}
\caption{ Pearson product-moment correlation coefficients area-based pulsality $P(1)$, tonality $T(1)$, and noisiness $N(1)$  and competing descriptors with MDS dimensions. The r-values for the competing descriptors are taken from \cite{Gygi_2007}, precise definition of these descriptors or pointers to them can be found there. Correlations printed in italics are negative.}
\label{tab:corrcoeffs}
\end{table}

The $P$ and $T$ descriptors were constructed to capture the organization of sounds along the first MDS dimension, while we constructed the $N$ descriptors to capture the organization along the second MDS dimension.  If we inspect figure~\ref{fig:logsumE_off_FS}.B we conclude that indeed $P(E)$ and $T(E)$ capture much of the organization along the first MDS dimension, this is reflected in the strong correlation between these descriptors, see table~\ref{tab:corrcoeffsenergybased}, and the first MDS dimension. In this scatter plot we see a clear, albeit imperfect,  separation of all three classes, however the visible separation is nonlinear and probably for that reason doesn't show up just as convincingly in the calculated correlation coefficients. Similarly we observe a separation of the three classes in panels C. and D. as well. 

If we inspect subfigure B. from figure~\ref{fig:surface_off_FS} we conclude that $P(1)$ and $T(1)$ still capture a large part of organization along the first MDS dimension, this is also visible in the strong correlation between $T(1)$ and the first MDS dimension. For $P(1)$ this correlation is however only moderate, see table~\ref{tab:corrcoeffs}. We also observe that the moderate correlation between $N(E)$ and the second MDS dimension has become weak in the case of $N(1)$. The effect of this reduction is visible in the scatter plots where the ability to distinguish between impact sounds and continuous sounds is completely lost. Harmonic sounds however can still be distinguished on the basis of these features. At least two factors contribute to this reduced separation of impact and continuous sounds. Firstly, pulsality is (per definition) high for a short interval and unless pulses are repeated rapidly and often, they occupy only a small part of the time-frequency-plane. Secondly, real-world pulses are often followed by reverberation, which shows low pulsality and high noisiness, and sometimes even increased tonality. So "reverberating" areas of the time-frequency plane contribute noisiness and perhaps tonality, but not pulsality. In the energy based version $P(E)$ the small areas with high pulsality are weighted with high energy, which ensures that $P(E)$ captures pulsality better than $P(1)$.

\section*{Discussion}

This paper substantiates the idea that sound perception and sound source recognition, as we propose, is based on sound production physics. The use of the features introduced in this paper is, however, not limited to the sound perception context. Instead it is a novel  technique to identify, describe, and quantify structure in time-series. In fact there are no obstacles to apply the techniques presented in this paper to other types of sequential data and arbitrary filter banks because of its explicit calibration steps in both the development of the mathematical framework as well as in the code used for its implementation. 

Because ~\cite{Gygi_2007} also did not know what (and if) categories would emerge from his perceptual similarity experiment, the dataset contains recordings which mix  in a single recording sounds from different perceptual categories, e.g. drums and cymbals. In addition unnatural silences are included, probably caused by audio editing during original construction of the sounds. In view of these dataset limitations, we have established a firm relation between our tract features and the perceptual dimensions. 
Our features are designed to be independent of discretization parameters. Previous work~\cite[]{Andringa2008,Andringa2010}, suffered from an implicit dependence on the discretization parameters: increasing the frequency or temporal resolution pulled more and more neighboring time-frequency points into the definition of the pulsality and tonality features (roughly analogous to $O_\ov$ and $O_\oh$, respectively). This included time-frequency points that were highly correlated with the center point. The correct behavior of the features depended therefore on resolution and in practice feature extraction only worked reliably in a single sound processing context. The previous system “suffered from” undocumented optimizations due to its evolutionary development in previous projects. The current definition and implementation solve these important limitations. 

We consider our work an example of computational auditory scene analysis (CASA). CASA broadly distinguishes two stages in sound perception: segmentation and grouping~\cite[]{Bregman1990,Wang2006,Szabo2016}. Our method offers a new segmentation method: “segments" that are dominated by pulses, tones or noises can be identified and characterized by thresholding of the tract features. These segments can be used directly in grouping algorithms or they can be used to select appropriate (local) analysis algorithms. Our work is based on a phenomelogical description of aspects of perception. It ignores most of the physiology of the auditory system. Instead it uses a more mathematical physics style of reasoning to constrain the choices leading up to our tract features. 

Applying the tract feature distributions in existing statistical tests is an attractive idea for machine learning purposes. A careful examination of conditions is required before, for example, the two-sided Kolomogorov-Smirnov test can be applied reliably. Both the time-frequency analysis and the feature extraction introduce correlations between measured values. This invalidates the assumption that each data point can be treated as an individual measurement. But the very notion of structured input invalidates it as well. Sparse sampling of the feature representations seems appropriate. But sparse sampling destroys time-translation invariance, which may complicate event recognition. For soundscape monitoring at long time scales, time-invariance is probably less relevant and sparse sampling might be viable. Nevertheless we think that calculating for example the Kolmogorov-Smirnov statistic between tract distributions for different sounds can be a viable first step in the construction of sound distance based classifiers.

We built tracts with two different orientations in the time-frequency plane: horizontal/temporal and vertical/spectral. The horizontal tracts were designed to respond strongly to tones and harmonic sounds with a slowly varying pitch. The vertical tracts were designed to respond strongly to pulses and onsets. As a bonus we observe that low values of our tract features are informative of structure in the perpendicular orientation. It is in principle also possible to use other orientations in the timefrequency-plane, which for example can facilitate the detection of frequency modulated signals. The  \verb!C++! code we used for this paper already allows for a larger variety of orientations. Also the notion of a temporal-spectral pattern can be generalized, which entails that it, in principle, is possible to detect harmonics or repetition of regularly repeating pulses.   

The confusions we found between continuous process and impact sounds (due to reverberation) in the second experiment suggest a role for rule based systems that argue about sound texture. This will, for example, allow us to identify a noise texture as reverberation if it is preceded by a pulse and decays exponentially. The noise texture can then, if suitably described, be attributed to either the preceding pulse or to unrelated other sources. We expect that this improves separation between impact and continuous process sounds. This helps to explain why listeners are usually insensitive to reverberation: reverberant energy is attributed to the preceding (driving) source and not treated as a separate independent signal. 

This is consistent with psychophysics and auditory scene analysis research that concluded that listeners segregate sound into physically coherent signal components, sometimes called partials or sound elements~\cite[]{Shamma_Micheyl_2011,Sussman_2005}. Listeners  (re)group these into the perceptual units that we become aware of (Bregman, 1995). These signal components, or what Winkler, Denham and Nelken refer to as 'feature-analyzed sound information' \cite[]{Winkler_Nelken_2009} must be treated as coherent perceptual units that can be grouped in a context of other signal components \cite[]{Sussman_2005}. In ASA experiments these units, conform the focus of our texture analysis, are typically tonal (as steady tones or as harmonics of speech), pulses or clicks, or noise bursts. The auditory system reliably estimates these signal components (partials, sound elements) from "begin to end" \cite[]{Andringa2002} and makes them available as the "feature-analyzed sound information". It is still unclear how to do this robustly and reliably. For example it complicates sound source analysis and recognition, especially in noise, if signal components originate from multiple sources because it starts as contribution of one source and ends as a contribution of another. Our texture analysis provides a new class of partials to work with, and hopefully an appropriate rule set can be designed to (re)group partials into units likely to belong to a single source.

\paragraph{Disclosure/Conflict-of-Interest Statement:}
The authors have founded a company SoundAppraisal B.V. to explore potential commercializations of the technology outlined in this paper. 

\paragraph{Author Contributions:}
Conceived the idea: RAJvE,TCA; Wrote the paper: RAJvE,TCA; Formalized the idea, wrote the feature extraction code: RAJvE; Analysed the results: RAJvE, TCA.   

\paragraph{Acknowledgement:}
The authors like to thank Coen Jonker for help in designing and coding the library for processing streamed data. The authors like to thank Bryan Gygi for kindly making available the data from the perceptual experiment. 

\paragraph{Funding:} 
Part of the research described in this paper was conducted as part of the Sensor City Sound project which was funded by the European Union, the European Regional Development Fund, the Dutch Ministry of Economic Affairs, and The Northern Netherlands Provinces Alliance, KOERS NOORD.


\end{document}